\documentclass[prl,twocolumn]{revtex4}

\usepackage[dvips]{graphicx}
\usepackage{amsmath}

\DeclareMathOperator{\erf}{erf}

\newcommand{\Dee}{\ensuremath{\mathcal{D}}}

\begin{document}

\title{Formal Solution of a Class of Reaction-Diffusion Models:   
Reduction to a Single-Particle Problem
}
\date{\today}
\author{Alan J.\ Bray$^{1,2}$}
\author{Satya N.\ Majumdar$^{2}$}
\author{Richard A.\ Blythe$^{1}$}
\affiliation{$^1$Department of Physics and Astronomy, University of 
Manchester, Manchester M13 9PL, U.K. \\
$^2$Laboratoire de Physique Quantique (UMR C5626 du CNRS), 
Universit\'e Paul Sabatier, 31062 Toulouse Cedex, France} 

\begin{abstract}

We consider the trapping reaction $A  + B \to B$ in space dimension $d
\le 2$. By formally eliminating  the $B$ particles from the problem we
derive  an effective  dynamics for  the $A$ particles  from  which the
survival probability of a given $A$ particle and the statistics of its
spatial fluctuations can  be calculated in a rather  general way.  The
method  can  be  extended  to the  study  of  annihilation/coalescence
reactions, $B+B \to 0$ or $B$, in $d=2$.
\end{abstract}

\maketitle


First-passage problems  involving more than  a few degrees  of freedom
are notoriously difficult to solve \cite{Sid,SatyaRev}. In this Letter
we  introduce  a  technique that  enables  one  to  solve a  class  of
first-passage  problems involving  an  infinite number  of degrees  of
freedom. For definiteness we develop  the method in the context of the
``trapping  reaction'', $A+B  \to B$,  but the  applications  are more
general, as emphasized in the latter part of the Letter. The main result  
of our approach is to reduce the problem to one described by a single 
degree of freedom whose late-time behavior can be extracted analytically. 

The  asymptotic   dynamics  of  the  trapping  reaction   has  been  a
long-standing  puzzle. The  main question  is how  the density  of $A$
particles decreases with time.  A related problem, much studied in the
context   of  chemical  kinetics   \cite{TW,BL}  is   the  two-species
annihilation reaction, $A+B \to  0$, with initial densities $\rho_A(0)
< \rho_B(0)$.   This is  equivalent to the  trapping reaction  at late
times, when  $\rho_A(t) \ll \rho_B(t)$ and  $\rho_B(t)$ is essentially
constant.  Again,  the standard problem  is to compute  the asymptotic
form  of the  $A$-particle density  $\rho_A(t)$ or,  equivalently, the
probability $Q(t)$ that a given  $A$ particle survives until time $t$.
Since the particles  do not interact with other  particles of the same
species, to compute $Q(t)$ it  suffices to consider a {\em single} $A$
particle  moving in  an infinite  sea  of $B$  particles with  density
$\rho$ ($=\rho_B$).

Traditional approaches to this type  of problem treat the $A$ particle
as an absorbing boundary for  the $B$ particles. Unfortunately, for an
arbitrary  $A$-particle  trajectory  this  absorbing-boundary  problem
cannot be solved.  In this Letter we introduce a different approach in
which  we  treat the  $A$  and  $B$ particles  as  if  they were  {\em
non-interacting}.   We exploit  the  initial condition  that each  $B$
particle  is randomly  located anywhere  in  the system  to show  that
certain `events', where a $B$ particle meets the $A$ particle {\em for
the   first   time}  (remember   that   we   are   treating  them   as
non-interacting,  so they  can meet  more  than once)  have a  Poisson
distribution, i.e.\  the probability $p_n$  that $n$ such  events have
occurred up  to time $t$ is  given by $p_n  = (\mu^n/n!)  \exp(-\mu)$,
where  the mean,  $\mu$, of  the distribution  is a  {\em functional},
$\mu[\vec{z}]$ of the trajectory  $\vec{z}(\tau)$, $0 \le \tau \le t$,
of   the  $A$   particle.    The  probability   that  the   trajectory
$\vec{z}(\tau)$  has  survived,  in  the  original  {\em  interacting}
problem,  is simply  $p_0[\vec{z}]  = \exp(-\mu[\vec{z}])$.   Finally,
$Q(t)$  is  obtained   by  averaging  $\exp(-\mu[\vec{z}])$  over  all
possible   $A$-particle    trajectories   $\vec{z}(\tau)$   with   the
appropriate       (Wiener)       measure,       $\exp[-(1/4D')\int_0^t
d\tau(d\vec{z}/d\tau)^2]$, where $D'$  is the $A$ particle's diffusion
constant. In  this way,  the $B$ particles  have been  {\em eliminated
from  the problem},  and one  has an  effective  $A$-particle dynamics
described by the Wiener measure and the functional $\mu[\vec{z}]$. The
final step,  that makes further  analytical progress possible,  is the
observation that  the path integral over  $\vec{z}(\tau)$ is dominated
at late times by a single $A$-particle trajectory.

The main  results of this approach  are: (i) the  trajectory where the
$A$ particle is stationary is proved to be the dominant trajectory and
determines  the  asymptotic  form   of  the  $A$  particle's  survival
probability \cite{BL}, $Q(t)  \sim \exp(-\lambda_d t^{d/2})$ for $d<2$
(with  a logarithmic  correction  in $d=2$),  where  $\lambda_d$ is  a
calculable  constant and  $d$  is the  dimensionality  of space;  (ii)
typical  {\em  fluctuations}  of surviving  $A$-particle  trajectories
around this  dominant path have variance $\langle  z^2(t) \rangle \sim
t^{(2-d)/2}$ for $d<2$ (iii) exact results are obtained for $Q(t)$ and
the form  of the dominant  path in a  system with a  {\em non-uniform}
initial  density  of $B$  particles;  (iv)  this  approach provides  a
powerful method  for calculating  first-passage properties for  a {\em
deterministically} moving  boundary $\vec{z}(t)$; (v)  the method also
provides a  formalism for calculating $Q(t)$ in  the highly nontrivial
situation where the $B$  particles themselves interact, e.g.\ $B+B \to
0$, at least in $d=2$  where the density correlations induced by these
reactions are negligible.

We begin by deriving the Poisson property that plays a central role in
the analysis. We consider a  finite volume, $V$, containing $N=\rho V$
$B$-particles (diffusion constant $D$), randomly distributed within it,
and a single $A$ particle (diffusion constant $D'$), initially located
at the origin. Let  $\vec{z}(t)$ be the $A$ particle's trajectory, and
let  $P({\vec{x}},t)$ be  the probability  that a  given $B$ particle,
starting at ${\vec{x}}$, has met the $A$ particle before time $t$. The
average of  this quantity over  the initial position,  ${\vec{x}}$, is
$(1/V)\int_V  dV\,P(\vec{x},t) = R(t)/V $, where $R(t)$ is an implicit 
functional  of   $\vec{z}(t)$.   The  probability  that  $n$  distinct
$B$ particles have  met the $A$ particle, averaged  over their initial
positions,  is $p_n(t)  =  \binom{N}{n} (R/V)^n (1-R/V)^{N-n}$. Taking  
the limit $N  \to \infty$, $V \to \infty$, with $\rho=N/V$ and $n$ held 
fixed,  one  recovers   the  Poisson  distribution, 
$p_n=(\mu^n/n!)\exp(-\mu)$, with $\mu = \rho R$. 

One can derive an equation for the functional $\mu[z]$ by calculating,
in two  ways, the probability  density to find  a $B$ particle  at the
point $\vec{z}(t)$ at time $t$. First, since the particles are treated
as  non-interacting, and  the $B$  particles start  in  a steady-state
configuration of  uniform density $\rho$, this  probability density is
just $\rho$. Secondly, from the Poisson property, the probability that
a $B$ particle (i.e.\ any $B$ particle) meets the $A$ particle for the
first time  in the time interval  $(t',t'+dt')$ is $\dot{\mu}(t')dt'$.
The probability density for such  a particle to subsequently arrive at
$\vec{z}(t)$  at  time  $t$  is  given  by  the  diffusion  propagator
$G(\vec{z}(t),t|\vec{z}(t'),t')      =      [4\pi      D(t-t')]^{-d/2}
\,\exp\{-[\vec{z}(t)-\vec{z}(t')]^2/4D(t-t')\}$.  Equating the results
from these two methods gives our fundamental equation,
\begin{equation}
\rho = \int_0^t dt'\,\dot{\mu}(t')\,G(\vec{z}(t),t|\vec{z}(t'),t')\ ,
\label{fundamental}
\end{equation}
which  is  an  implicit  equation for  the  functional  $\mu[\vec{z}]$
(noting  that  $\mu(t=0)=0$,  since   no  $B$ particle  can  meet  the
$A$ particle in  zero time). Finally, 
$Q(t)=\langle\exp(-\mu[\vec{z}])\rangle_z$, where the average is  
over  all paths $\vec{z}(t)$ weighted with the Wiener measure.

As a first  application of this equation we  prove that the trajectory
$\vec{z}=0$  is  the dominant  path,  i.e.\  that it  gives  the
smallest possible value of $\mu[\vec{z}]$ for all $t$. This  function,
$\mu_0(t)$, satisfies Eq.\ (\ref{fundamental}) with $\vec{z}=0$:
\begin{equation}
\rho = \int_0^t dt'\,\dot{\mu}_0(t')[4\pi D(t-t')]^{-d/2}\ .
\label{mu_0}
\end{equation}
By  inspection, $\mu_0(t)$ must  have the  form $\mu_0(t)  = \lambda_d
t^{d/2}$ (for $d<2$), in order that the right-hand side be independent  
of $t$. Substituting  this form in (\ref{mu_0}), and evaluating the 
integral, gives 
\begin{equation}
\lambda_d = \rho\left(\frac{2}{\pi d}\right)\,
\sin\left(\frac{\pi d}{2}\right)\,\left(4\pi D\right)^{d/2},\ \ \  d<2,
\label{lambda}
\end{equation}
while for $d=2$ one finds,  for $t \to \infty$, $\mu_0(t) \to 4\pi\rho
Dt/\ln   t$   \cite{d=2}.   The  corresponding   A-particle   survival
probability  is $Q_0(t)  =  \exp[-\mu_0(t)]$. This  simple  case of  a
static  $A$  particle is  sometimes  called  the `target  annihilation
problem', and our method reproduces the known results for that problem
\cite{target} in a very simple way. To prove that $\vec{z}(t)=0$ gives
the global minimum of $\mu[\vec{z}]$ we write $\mu = \mu_0 + \mu_1$ in
(\ref{fundamental}).  This  equation can then be  rearranged, with the
help of Laplace transform techniques, to give an implicit equation for
$\mu_1[\vec{z}]$:
\begin{eqnarray}
\mu_1[\vec{z}] & = & \frac{\sin(\pi d/2)}{\pi} \int_0^t
\frac{dt_1}{(t-t_1)^{(2-d)/2}}  \nonumber  \\  
&& \times  \int_0^{t_1}
\frac{dt_2}{(t_1-t_2)^{d/2}}\,\dot{\mu}(t_2) K(t_1,t_2),
\label{ubound}
\end{eqnarray}
where 
$K(t_1,t_2)= 1 - \exp\{-[\vec{z}(t_1)-\vec{z}(t_2)]^2/4D(t_1-t_2)\}$.  

Eq.\ (\ref{ubound})  is `implicit' because  the full $\mu$  appears on
the right-hand side.  Now note  that $K(t_1,t_2) \ge 0$ and $\dot{\mu}
\ge 0$ (because $\mu(t)$ is the mean number of different $B$ particles
that  have  met  the  $A$  particle  up  to  time  $t$  --  clearly  a
non-decreasing  function).  Therefore $\mu_1[\vec{z}]  \ge 0$  for all
paths $\vec{z}(t)$, with equality  when $\vec{z}(t)=0$ for all $t$. It
follows that $Q(t) \equiv \langle \exp(-\mu_0-\mu_1) \rangle_{\vec{z}}
\le \exp[-\mu_0(t)]$.  This rigorous  upper bound for $Q(t)$, combined
with the  identical rigorous lower bound derived  in \cite{BB}, proves
that the  asymptotic form of  $Q(t)$ is the  same as for  the `target'
problem, where the $A$ particle is  stationary, for all $d \le 2$. The
interpretation of this result is  that, since $\mu$ is {\em large} for
$t \to  \infty$ ($\mu \sim  t^{d/2}$), the path-integral for $Q(t)$ is
dominated by the  path that minimizes $\mu$, i.e.\  we are essentially
evaluating the path integral by the method of steepest descents. Small
fluctuations around  the dominant path will  determine the corrections
to the asymptotic form. 

We  next  compute  the  probability  distribution,  $P(z,t)$,  of  the
position $z$ of the $A$ particle at time $t$,  given that it survives.
Numerical studies  \cite{MG} suggest  that, in $d=1$,  $\langle z^2(t)
\rangle^{1/2}  \sim t^\phi$,  with  $\phi =  0.25-0.3$, while  similar
studies  in $d=2$ are  inconclusive.  Our  methods give  $\phi=1/4$ in
$d=1$ and  $\phi=(2-d)/4$ for all  $d<2$.  The technique is  to expand
$\mu_1[\vec{z}]$,  given by  Eq.\  (\ref{ubound}), to  order $z^2$  to
compute the variance of the  Gaussian fluctuations around the dominant 
trajectory $z(t)=0$.  To this order one can replace  $\mu(t_2)$ on the 
right-hand side by $\mu_0(t_2) = \lambda_d t_2^{d/2}$,  and expand the 
function $K(t_1,t_2)$ to order $z^2$. Specializing to $d=1$, the result 
is, at time $t$,
\begin{equation}
\mu_1[z] =  \frac{\lambda_1}{8\pi D}\int_0^t \frac{dt_1}{\sqrt{t-t_1}}
\int_0^{t_1}dt_2\frac{[z(t_1)-z(t_2)]^2}{\sqrt{t_2}(t_1-t_2)^{3/2}}.
\end{equation} 

The probability  distribution for $z$ at  time $t$ is given,  up to an
overall normalization, by the path integral
\begin{equation}
P(z,t)  =   \int  \Dee  z(t)   \exp\left(-\frac{1}{4D'}\int_0^t  d\tau
\dot{z}^2(\tau) - \mu_1[z]\right),
\end{equation}
where the integral is over all paths satisfying the boundary conditions
$z(0)=0$, $z(t)=z$.

The path integral has the form $\int \Dee z(t) \exp(-S[z])$, where the
`action'  $S[z]  = S_T[x]  +  S_V[z]$  is  a quadratic  functional  of
$z(\tau)$, where $S_T[z]=  (1/4D')\int_0^t d\tau \dot{z}^2(\tau)$, and
$S_V[z]   =  \mu_1[z]$.    Since  the   integrand  is   Gaussian,  the
$z$-dependence  of  the  path  integral  is exactly  captured  by  the
extremal path connecting $z(0)=0$ and $z(t)=z$.  Although we have been
unable to find this path analytically, power counting on the two terms
suggests $S_V \sim \lambda_1z^2/D\sqrt{t}$  and $S_T \sim z^2/D't$, so
that  $S_V$   dominates  at  large   $t$.  A  more   careful  analysis
\cite{details} confirms  the dominance  of $S_V$ and  the form  $S_V =
c_V\lambda_1z^2/D\sqrt{t}$,  where  $c_V$   is  a  pure  number.   The
probability   weight  for   the  fluctuations   $z(t)$   of  surviving
trajectories  is  therefore Gaussian,  with  variance $\langle  z^2(t)
\rangle = (D/2c_V\lambda_1)\sqrt{t}$ for  large $t$, with the value of
$c_V$  determined by  the  extremal  path. Note  that  this result  is
independent of the $A$-particle diffusion constant $D'$. If $D'=0$, of
course, there  are no  fluctuations, but for  any $D'>0$  the variance
becomes independent of $D'$ at large enough $t$. Also, since typically
$z  \sim t^{1/4}$,  the next  term in  the expansion  of  the function
$K(t_1,t_2)$  in  (\ref{ubound}) is  of  relative  order $z^2/Dt  \sim
t^{-1/2}$ so it  is negligible at large $t$.   The distribution of $z$
is therefore exactly Gaussian, at  least in the `scaling limit' $z \to
\infty$,  $t \to  \infty$, with  $z/t^{1/4}$ fixed. Similar arguments
\cite{details}  give  the  generalization $\langle z^2(t) \rangle \sim
(D/\lambda_d) t^{(2-d)/2}$ for $d<2$. 

We turn now to a related problem with a non-trivial dominant path that
can be  exactly determined.   Consider, in $d=1$,  a system  where the
density of $B$  particles at $t=0$ has different  values, $\rho_L$ and
$\rho_R$, to the left and right of the $A$ particle. The derivation of
an equation for  $\mu[z]$ proceeds exactly as before,  except that the
probability density  to find a $B$  particle at the point  $z$ at time
$t$ in  the noninteracting system, which  appears on the  left of Eq.\
(\ref{fundamental}), has to be recalculated. In terms of the diffusion
propagator  $G$   introduced  earlier,  this   probability  is  (quite
generally) $P_B(z,t) = \int_{-\infty}^\infty dx\, \rho(x)\,G(z,t|x,0)$
where  $\rho(x)$  is  the  initial $B$-particle  density  at  position
$x$. When  $\rho(x)=\rho$, a  constant, one finds  $P_B(z,t)=\rho$, as
before. When  $\rho(x) = \rho_{L}$  for $x<0$ and $\rho_R$  for $x>0$,
the generalized version of (\ref{fundamental}) becomes
\begin{equation}
\rho\left[1-\Delta\erf\left(\frac{z(t)}{\sqrt{4Dt}}\right)\right] 
= \int_0^t dt'\,\dot{\mu}(t')\,G(z(t),t|z(t'),t'),
\label{assym}
\end{equation}
where $\rho = (\rho_L+\rho_R)/2$ is now the {\em mean} density, 
$\Delta = (\rho_L-\rho_R)/(\rho_L + \rho_R)$ is a measure of the 
left-right asymmetry, and $\erf(x)$ is the error function. 

Physical intuition suggests that,  because of the asymmetry, surviving
$A$-particle trajectories  will tend to  be those that drift  into the
region (the  right, say) where  the $B$-particle density  is initially
smaller. Upper  and lower bounds have been  derived earlier \cite{BB2}
for the asymptotics of  the $A$-particle survival probability, $Q(t)$,
which  show  that  it  has  the  asymptotic  form  $Q(t)  \sim  \exp[-
g(\Delta)\lambda_1\sqrt{t}]$, where $g(0)=1$  for consistency with the
symmetric  case,  $\rho_L=\rho_R$. This  form  for  $Q(t)$ shows  that
$\mu[z]$  for the optimal  path has  the time-dependence  $\mu \propto
\sqrt{t}$.   Both   sides  of  (\ref{assym})  can   then  be  rendered
time-independent by the choice $z(\tau) = \alpha\sqrt{4D\tau}$ for all
$\tau \le t$,  where $\alpha$ is a constant to  be determined.  A more
detailed  analysis  \cite{details} shows  that  the  dominant path  is
indeed of this form.

Putting  $\mu(t) = g(\Delta)\lambda_1\sqrt{t}$ and
$z(\tau)=\alpha\sqrt{4D\tau}$, in (\ref{assym}), and evaluating the
integral on the right-hand side gives
\begin{equation}
g(\Delta)   =  \frac{\exp(-\alpha^2)}{1   -  \erf^2(\alpha)}\,   
[1  - \Delta\erf(\alpha)].
\label{g}
\end{equation}  
The  final step is  to minimize  the right-hand  side with  respect to
$\alpha$ to  obtain the  optimal path. This  can be  done numerically.
The resulting $g(\Delta)$ is shown for  $\Delta \ge 0$ as the inset in
Figure  1 (note that,  by symmetry,  $g(\Delta)$  is symmetric  around
$\Delta=0$). It clearly satisfies the bounds $1-|\Delta| \le g(\Delta)
\le 1$ derived in \cite{BB2}. 
\begin{figure}
\includegraphics[width=\linewidth]{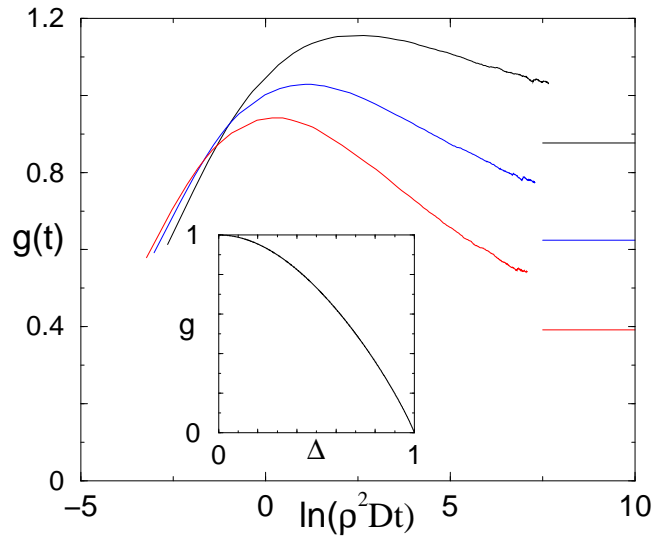}
\vspace*{-0.8cm}
\caption{\label{figure}   Time-dependence   of   $g(t)   \equiv   -\ln
Q(t)/\lambda_1\sqrt{t}$ for density  asymmetries $\Delta = 1/3$, $3/5$
and $7/9$ (top  to bottom).  The horizontal lines  show the asymptotic
value of  $g$ in  each case.  while  the inset  shows this value  as a
function of $\Delta$.}
\end{figure}

In the main part of Fig.\  1, numerical results for $g(t) = -\ln Q(t)/
\lambda_1\sqrt{t}$, obtained using the algorithm of ref.\cite{MG}, are
displayed, with $D=1/2$ and $\rho_L=0.5$ in all cases, while $\rho_R =
0.25, 0.125,  0.0625$ (top to  bottom on the right),  corresponding to
$\Delta =  1/3, 3/5,  7/9$ respectively. The  horizontal lines  on the
right  show the  asymptotic values  obtained  from the  inset for  the
corresponding values  of $\Delta$. The slow approach  to asymptopia is
similar  to   that  observed  \cite{MG,BB2}  in   the  symmetric  case
($\Delta=0$).

As a bonus,  the same calculation solves the  first-passage problem of
$B$ particles with  density $\rho_L$ for $x<0$ and  $\rho_R$ for $x>0$
moving  in  the  presence  of  a  deterministically  moving  absorbing
boundary located at $x(t) = \alpha\sqrt{4Dt}$. The probability that no
particle has reached  the boundary up to time $t$  has the simple form
$Q(t)=\exp[-g(\Delta)\lambda_1\sqrt{t}]$,  with  $g(\Delta)$ given  by
(\ref{g}).   We are  not  aware of  any  other way  of obtaining  this
result. Extensions to deterministically moving absorbing boundaries in
dimension $d>1$ are also possible \cite{details}.

In  the  final part  of  this  Letter, we  apply  this  approach to  a
nontrivial problem  with $d=2$. Consider  the annihilation/coalescence
reaction $B + B \to 0$  with probability $1/(q-1)$ and $B+B\to B$ with
probability $(q-2)/(q-1)$.   The density of $B$ particles  is known to
decay as $\rho(t) = a_d  [(q-1)/q] (Dt)^{-d/2}$ for $d<2$, where $a_d$
is a universal constant \cite{Lee,Cardy} equal to $1/2\pi\epsilon$ for
$d \to 2$, with $\epsilon=2-d$. For $d=2$, a logarithmic correction is
obtained,  $\rho(t) \sim \ln  t/t$. Now  suppose that  one of  the $B$
particles  is tagged  and relabeled  an $A$  particle,  with diffusion
constant $D'$.  We consider  the probability (the `walker persistence'
probability \cite{WP}), $Q(t)$, that the  $A$ particle has not met any
$B$  particle  up  to time  $t$.   The  limit  $d  \to 2$  provides  a
simplification because it is  the borderline dimension above which the
rate  equation  approach,  $d\rho/dt  \propto  -\rho^2$,  which  gives
$\rho(t)  \propto  1/t$,  is  qualitatively  correct  because  density
fluctuations can  be ignored \cite{Lee,Cardy}.  Eq.(\ref{fundamental})
is readily adapted to this case.  As before, we treat the $A$ particle
as  if  it  does  not  interact  with the  $B$  particles,  while  the
interactions of the  $B$ particles with each other  give rise to their
decreasing density, $\rho(t)$. The Poisson distribution for the number
of first  crossings of the $A$  particle by $B$  particles still holds
for this system. The left-hand side of Eq.\ (\ref{fundamental}), i.e.\
the probability density to find a  $B$ particle at the point $z(t)$ at
time  $t$,  becomes  $\rho(t)$,  while  on  the  right-hand  side  the
propagator  $G(z(t),t|z(t'),t')$  has to  be  multiplied  by a  factor
$\rho(t)/\rho(t')$,  being the  probability  of a  given $B$  particle
surviving till  time $t$  given that it  survives till time  $t'$. The
required generalization of (\ref{fundamental}) then reads
\begin{equation}
1 = \int_0^t \frac{dt'}{\rho(t')}\,\dot{\mu}(t')\,
G(\vec{z}(t),t|\vec{z}(t'),t').
\label{fund2d}
\end{equation}

It is convenient to approach the  limit $d \to 2$ from below. Consider
first the case $D'=0$, for which $Q(t)$ becomes the `site-persistence'
probability, i.e.\  the probability that  a given point in  space (the
location  of  the  $A$ particle)  has  not  been  visited by  any  $B$
particle.  The static $A$ particle corresponds to $z(\tau)=0$, for all
$\tau$,  and   with  $\rho(t') = a_d [(q-1)/q] (Dt')^{-d/2}$ for large  
$t$, (\ref{fund2d}) becomes
\begin{equation}
(4\pi)^{d/2}a_d(q-1)/q   =  \int_0^t  dt'\,\dot{\mu}(t')\,{t'}^{d/2}\,
(t-t')^{-d/2}
\label{2d}
\end{equation}
for large $t$.  In order  that the right-hand side be time-independent
for large  $t$, $\mu(t)$  must have the  asymptotic form  $\mu(t) \sim
\theta \ln  t$.  Inserting this  form into (\ref{2d}),  and evaluating
the       integral      gives      $\theta       =      2^d\pi^{d/2-1}
\sin(\pi\epsilon/2)a_d(q-1)/q$.   Taking the  limit $\epsilon  \to 0$,
using $a_d =  1/2\pi\epsilon$ in this limit, gives  $\theta = (q-1)/q$
for $d=2$.   Finally, $Q(t) =  \exp[-\mu(t)] \sim t^{-\theta}$  for $t
\to \infty$.   The result $\theta=(q-1)/q$ in $d=2$ agrees with that  
obtained by Cardy \cite{Cardy} using field-theoretic methods.
    
If one now considers the  case $D'>0$, i.e.\ a diffusing $A$ particle,
one sees  immediately that for $D'=D$  and $q=2$ (so that  $B+B \to 0$
always), the $A$  particle is equivalent to another  $B$ particle, and
$Q(t) = \rho(t)$, the density,  since every surviving particle has not
met  any other  particle.  Hence  $Q(t) =  \rho(t) \sim  \ln  t/t$ for
$D'=D$ and  $q=2$. This suggests  that, for general $D'$,  $Q(t)$ will
decay as $t^{-\theta}$ where $\theta$  is a nontrivial function of
$D'/D$.

The calculation of $\theta$ can readily be extended to $D' > 0$ within
the present formalism, using  a power-series expansion in $D'/D$.  The
method  is  to use  a  cumulant expansion  to  write  $Q(t) =  \langle
\exp(-\mu) \rangle_z  = \exp(-\langle \mu \rangle_z  + \{\langle \mu^2
\rangle_z - \langle \mu \rangle_z^2\}/2! + \cdots)$. To first order in
$D'$,  one needs  only the  first cumulant.  The result  is  $\theta =
(1+D'/D)(q-1)/q$.  This  class of problems,  where $Q(t)$ decays  as a
power-law  instead of  a stretched  exponential, marks  the borderline
where the path-integral for $Q(t)$  is no longer dominated by a single
path (and small fluctuations about  it), giving a leading large-$t$ 
form independent  of $D'$,  but has  to be  evaluated exactly,  with 
results that depend  on  $D'$  even for $t \to \infty$.  Full  details  
of  this calculation, together with results to  higher order in $D'/D$, 
will be presented elsewhere.

In conclusion, we have introduced an analytic approach to a class 
of reaction-diffusion models that reduces them to one-particle systems. 
We hope to use this method in future to address, {\em inter alia}, 
the vexing problem of the very slow approach to asymptopia in the 
trapping reaction.   

RAB thanks the EPSRC for financial support under grant GR/R53197.


\begin{references}

\bibitem{Sid} 
S. Redner, {\em A guide to first-passage processes} (CUP, Cambridge, 2001).

\bibitem{SatyaRev}
S. N. Majumdar, Curr.\ Sci.\ {\bf 77}, 370 (1999).

\bibitem{TW}
D. Toussaint and F. Wilczek, J. Chem.\ Phys.\ {\bf 78}, 2642 (1983).

\bibitem{BL}
M. Bramson and J. L. Lebowitz, Phys.\ Rev.\ Lett.\ {\bf 61}, 2397 (1988).

\bibitem{d=2}
To obtain sensible results for $d \ge 2$ the A particle has to be given 
a nonzero radius $r_0$, which, for $d=2$, only enters the final result 
as a time scale, $t_0=r_0^2/4D$, in the logarithm:  
$\mu_0(t) = 4\pi\rho Dt/\ln(t/t_0)$. For $d>2$ the present method cannot 
be used as it stands. 

\bibitem{target}
A. Blumen, G. Zumofen, and J. Klafter, Phys.\ Rev.\ B {\bf 30}, 5379 
(1984). 

\bibitem{BB}
A. J. Bray and R. A. Blythe, Phys.\ Rev.\ Lett.\ {\bf 89}, 150601 (2002).  

\bibitem{MG}
V. Mehra and P. Grassberger, Phys.\ Rev.\ E {\bf 65}, 050101 (2002). 

\bibitem{details}
The details will be presented elsewhere. 

\bibitem{BB2}
R. A. Blythe and A. J. Bray, cond-mat/0209353, submitted to Phys.\ Rev.
\ E.

\bibitem{Lee}
B. P. Lee, J. Phys.\ A {\bf 27}, 2633 (1994).

\bibitem{Cardy}
J. Cardy, J. Phys.\ A {\bf 28}, L19 (1995).

\bibitem{WP}
C. Monthus, Phys.\ Rev.\ E {\bf 54}, 4844 (1996); S. N. Majumdar 
and S. J. Cornell, Phys.\ Rev.\ E {\bf 57}, 3757 (1998); S. J. 
O'Donoghue and A. J. Bray, Phys.\ Rev. E {\bf 65}, 051114 (2002). 

\end{references}

\end{document}